\documentstyle[emulateapj,flushrt,epsf,rotate]{article}

\long\def\jumpover#1{{}}

\def\approxgt{\,\raise2pt \hbox{$>$}\kern-8pt\lower2.pt\hbox{$\sim$}\,}
\def\approxlt{\,\raise2pt \hbox{$<$}\kern-8pt\lower2.pt\hbox{$\sim$}\,}
\def \th{\thinspace}

\def \ni{\noindent}
\def \Teff{{$T_{\rm {ef\!f}} $}}
\def \Mo{{$M_\odot $}}
\def \Lo{{$L_\odot $}}
\def \at{{\rm\char'100}}

\def \eg{{{\it e.g.},\ }}
\def \etal{{\it et al.\ }}

\def \cf{{\it cf.\ }}

\def \ie{{{\it i.e.},\ }}

\def \viz{{\it viz.\ }}
\def \vs{{\it vs.\ }}

\def\Log{{\mathrm Log}}
\def\LogL{{Log$\th L$}}
\def \LTeff{{Log$\th T_{{\rm ef\!f}} $}}
\def\LogP{{Log$\th P$}}

\def\rn{\noindent\parshape 2 0truecm 8.8truecm 0.3truecm 8.5truecm}

\begin{document}

\submitted{ASTROPHYSICAL JOURNAL, revised version, in press}

\title{RR Lyrae -- Theory \lowercase{vs} Observation}
\author{Zolt\'an Koll\'ath$^{1}$,
J. Robert Buchler$^{2}$
 \& Michael Feuchtinger$^{2}$
}
 \begin{abstract}

The luminosities, effective temperatures and metallicities that are derived
empirically by Kov\'acs and Jurcsik from the light curves of a large number of
globular cluster and field RRab and RRc stars are compared to theoretical RR Lyrae
models.  The strong luminosity dependence of the empirical blue and red edges
(\LogL\ \vs \LTeff\ diagram) is in disagreement with that of both radiative and
convective models.  A reexamination of the theoretical uncertainties in the
modelling leads us to conclude that the disagreement appears irreconcilable.

\end{abstract}


{\bigskip
        {\footnotesize
\noindent $^1$Konkoly Observatory, Budapest, HUNGARY; kollath\at
konkoly.hu  \\
\noindent $^2$Physics Department, University of Florida, Gainesville, FL, USA;
buchler\at phys.ufl.edu, fm\at phys.ufl.edu
}}

 \section{Introduction}

The recent work of Kov\'acs \& Jurcsik (Kov\'acs \& Jurcsik 1996 [KJ96], 1997
[KJ97], Jurcsik 1998 [J98]) proposes an almost purely empirical method of
extracting the absolute magnitudes, colors and effective temperatures (\Teff)
and metallicities directly from the observed periods and light curves of RR
Lyrae stars.  The only theoretical input appears in the transformation from
M$_V$ and V -- K to L and \Teff\ via Kurucz's static model atmospheres.  The
very empirical nature and potential usefulness of the approach has attracted a
great deal of attention from observers.

The most recent work (J98) compiles and analyzes a sizeable set of
observational data of RR Lyrae cluster variables and RR Lyrae field stars.
This study thus includes RR Lyrae stars with metallicities ranging from
Z=0.00001 to almost solar Z=0.020.  The end products of her analysis that we
are most concerned about here are \LogL-\LTeff\ and \LogL-\LogP\ plots, and the
byproducts which are relations between luminosity L, mass M and metallicity Z.

The J98 \LogL-\LTeff\ and \LogL-\LogP\ data are shown in Figure~1 as small dots,
circles for RRab and triangles for RRc stars.  The figures indicate well
defined slopes for the fundamental and overtone blue edges and red edges, all
four of which have approximately the same values, although there may be a
slight broadening of the instability strip with luminosity.  Hereafter we call
this slope $\Xi$ ($=\Delta$~\LogL / $\Delta$~\LTeff).

J98 pointed out that no model calculations can explain the strong dependence of
the temperature on the luminosity.  In \S2 we first confirm that indeed the
slope $\Xi$ of this empirical \LogL\ \vs \LTeff\ diagram is in irreconcilable
disagreement with that of radiative models.  Next, in \S3 we show that the
inclusion of turbulent convection in the models only shifts the \LogL\ \vs
\LTeff\ line, but does not change the slope.  The discrepancy is therefore very
basic because it is already inherent in models that are as fundamental as
purely radiative ones.  In \S4 we reexamine the uncertainties of both radiative
and turbulent model calculations and conclude that the theoretical $\Xi$ is
very robust and cannot be changed very much without introducing new physics.
In \S5 we discuss the theory that goes into the empirical relations of Jurcsik,
in particular the influence of the use of the static Kurucz color --
temperature transformation.  We conclude in \S6.

 \section{Radiative Models of RR Lyrae Stars}

In the following we examine how well RR Lyrae models agree with the J98 data.
We first examine radiative models, and thus limit ourselves to the vicinity of
the blue edges.

\begin{figure*}
\vspace{0cm}
\centerline{{\vbox{\epsfxsize=9cm\epsfbox{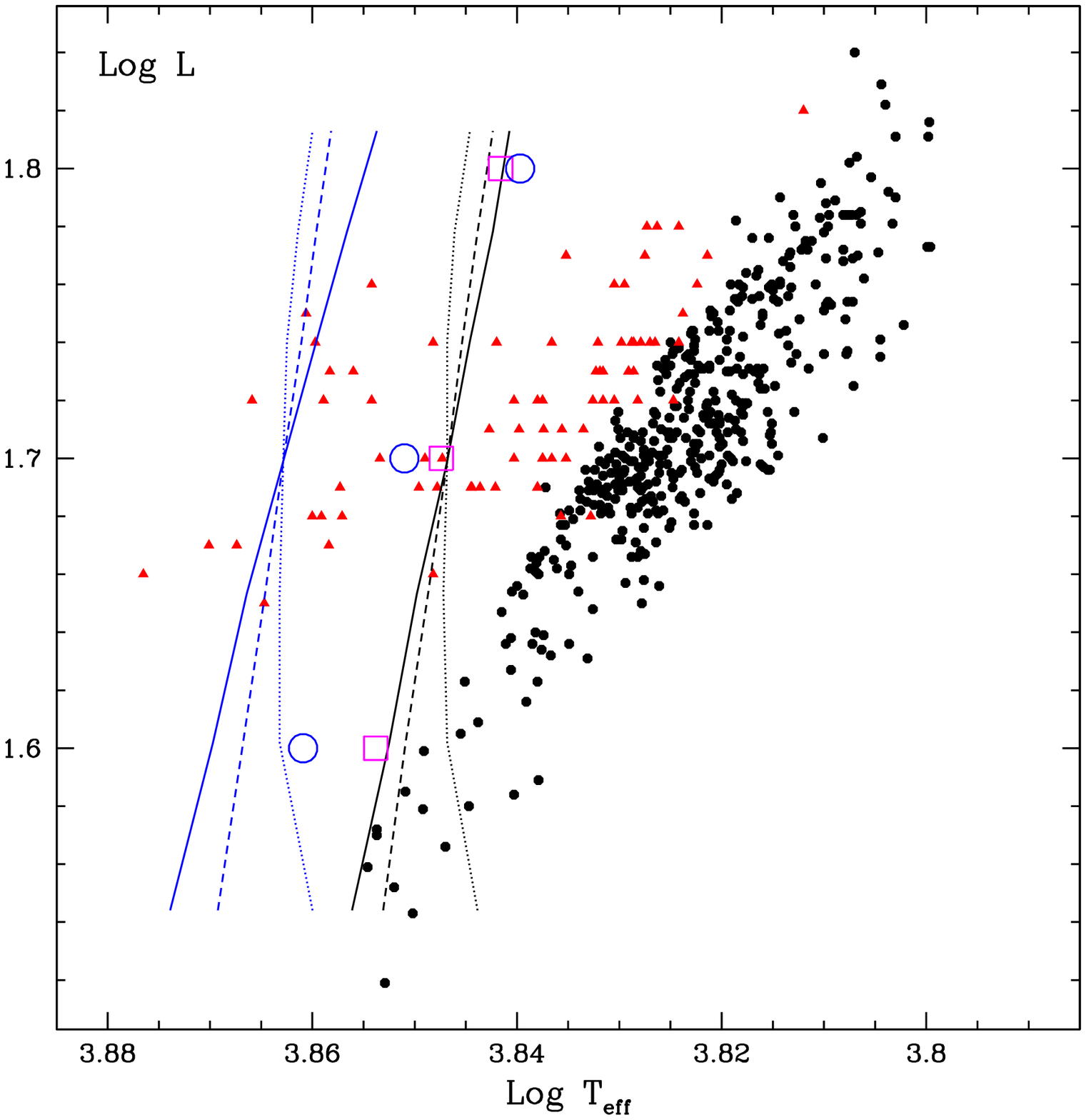}}}
            {\vbox{\epsfxsize=9cm\epsfbox{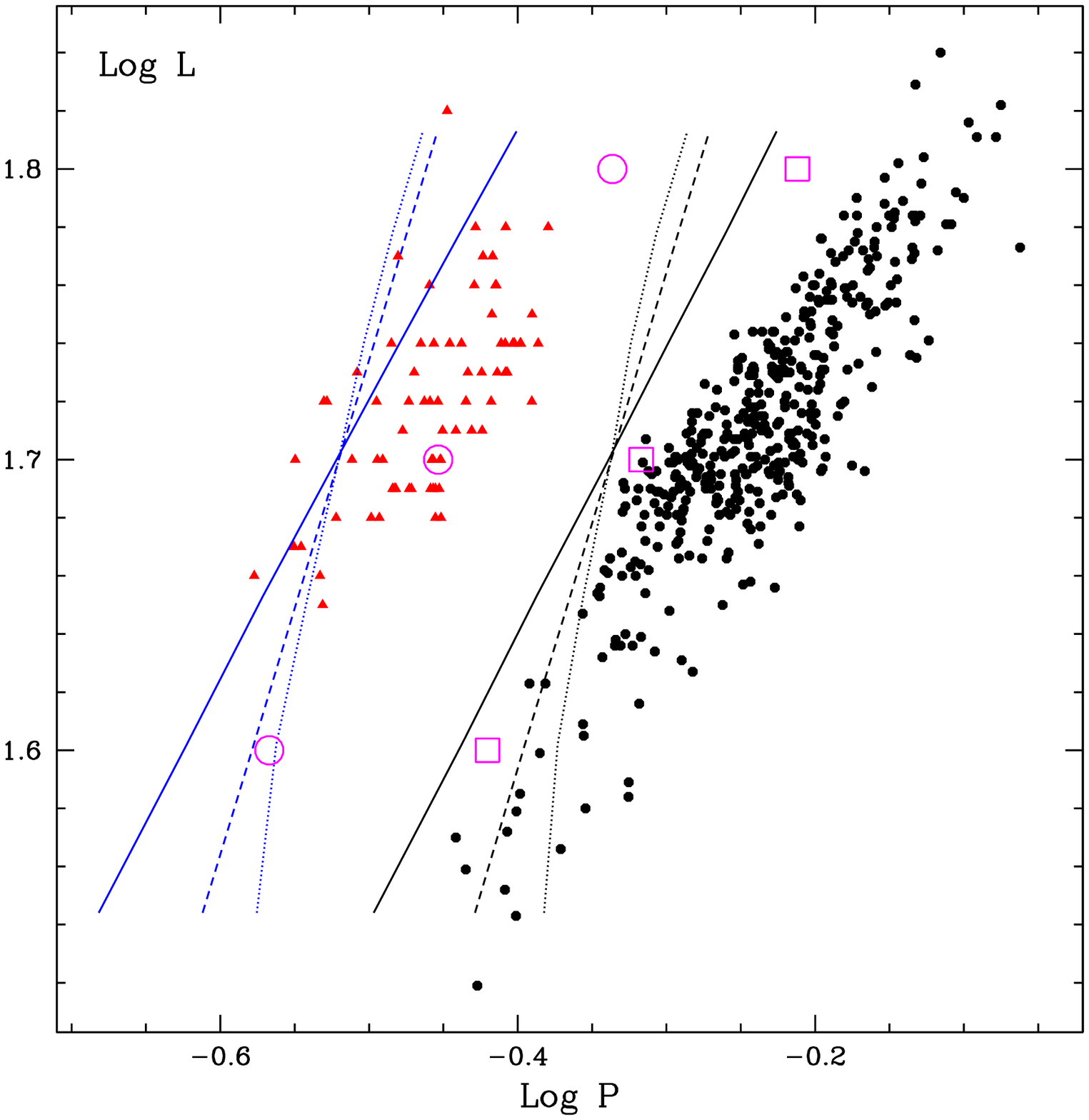}}}}
\vspace{0.2cm}
\noindent{\small
 Fig.~1 :
Left: \LogL-\LTeff\ plot:
 RRab (filled circles) and RRc (triangles) stars from Jurcsik 
([J98] and priv. comm.);
 Theoretical fundamental and first overtone blue edges for radiative models:
{\it solid lines}: fixed mass and Z;  
{\it dotted lines} have M and Z  from Jurcsik's relation; 
{\it dashed lines} have M from Jurcsik with fixed Z=0.001; 
Tuggle \& Iben's overtone and fundamental blue edges are reported as open
circles and squares, respectively.
Right:  Same data in a \LogL-\LogP\ plot.
 }
 \label{fig1}
\end{figure*}

Figure~1 shows a \LogL-\LTeff\ plot on the left.  On these we superpose the
linear fundamental and first overtone blue edges of radiative RR Lyrae models.
We recall here that among the linear edges of the instability strip only the
overtone linear blue edge and the fundamental linear red edge coincide with the
observable edges of the instability strip.  Nonlinear dynamical effects shift
the observable fundamental blue edge and the overtone red edge to the red and
to the blue, respectively, compared to their corresponding linear edges, by up
to several 100\th K (\eg \cf Buchler 2000).  Strictly speaking, for radiative
models, it is only the linear overtone blue edge that is relevant, but the
linear fundamental blue edge can give a rough indication of the slope $\Xi$.
Furthermore, the red edges are determined by convection, and radiative red
edges are not relevant and are therefore not shown in Fig.~1.

The solid lines have been computed for models with M=0.65, X=0.75 and Z=0.001.
The dotted lines represent models with M and Z from Jurcsik's \{L, M, Z\}
relations, the dashed  lines those with M from Jurcsik, but with Z=0.001.
Finally, the large open circles are models from Tuggle and Iben (1972) for
M=0.6 and Z=0.001 (with the old Los Alamos opacities).  They have essentially
the same $\Xi$ as ours which are computed with the OPAL opacities (Iglesias \&
Rogers 1996) merged with the low temperature Alexander-Ferguson (1994) opacities.

Again, because we are computing radiative models we do not expect the location
of the blue edge to be in perfect agreement with the edge of the instability
strip, but we note a very large discrepancy in the slope $\Xi$.

Figure~1 also displays a \LogL\ -- \LogP\ plot on the right.  Only the slope of
the constant mass models is in almost acceptable agreement with the J98 data,
but those calculated with the J98 $\{M, L, Z\}$ are in strong disagreement.

In Figures~2 we display again a \LogL\ -- \LTeff\ plot which shows the effect of
composition on the location of the linear blue edges.  The (radiative) models
were computed with M=0.65. Here the solid lines have X=0.75, Z=0.001, the dotted
lines have X=0.75, Z=0.004 and the dashed lines have X=0.70, Z=0.001.

The location of the blue edge displays very little sensitivity to either
metallicity Z or helium content Y, within a reasonable range of values.
Adjusting the helium content or metallicity does not provide a resolution of
the slope discrepancy.  We  describe further tests in \S4.

We conclude this section by noting that purely radiative RR Lyrae models are in
severe disagreement with Jurcsik's \th \LogL-\LTeff\ data.

 \section{Convective  Models of RR Lyrae Stars}

The addition of time-dependent turbulent convection in the models has led to
some spectacular successes compared to radiative models, in particular, in
finally predicting double mode behavior both in RR Lyrae (Feuchtinger 1998)
and in Cepheids (Koll\'ath, Beaulieu, Buchler \& Yecko 1998).  One would hope
therefore that the inclusion of turbulent convection might also remove the
discrepancy between observations and theory described in the previous section.

Yet, in that respect all of our modelling efforts with turbulent convection
have proved in vain, despite the flexibility afforded by the 8 free order unity
($\alpha$) parameters that the turbulent convective equations contain (\eg
Yecko, Koll\'ath \& Buchler 1997, Buchler\& Koll\'ath 2000).  We find that
turbulent convection can shift the blue edges, but cannot produce the
differential effect with respect to luminosity that is required to give the
right slope $\Xi$ in the \LogL\ \vs \LTeff\ plots.  In Figure~3 we show the
results for three different combinations of these parameters.

We note that different codes and slightly different recipes for convection
give essentially the same theoretical slopes.  For example, even though no
linear models are computed, the nonlinear hydrodynamical models of Bono \etal
(1997) indicate a similarly steeper slope than that of J98.

 \section{Discussion of the Modeling Assumptions}

 In this section we discuss the effects of several of the approximations that
 are inherent in the numerical modeling to see how robust the theoretical
 \LogL-\LTeff\ slope is.

 \subsection{Equation of State}

Our code uses as equation of state a simple iteration of the Saha equations
(e.g. Stellingwerf 1982).  This equation of state is very similar to the OPAL
(Rogers \etal 1996) and the MHD (D\"appen \etal 1988) equations of state and we
do not believe that the tiny differences can be responsible for the discrepancy
between the models and the J98 data.

\vfill\eject

 \subsection{Radiative Transport}

Our radiative code uses a standard equilibrium diffusion approximation for
the radiative transport, \ie

\begin{equation}
 L = (4\pi R^2)^2 {c\over 3\kappa} {\partial\over \partial m} aT^4
\label{L_eq}
\end{equation}

\ni in which one uses for the opacity $\kappa$ the Rosseland mean (\eg Mihalas
\& Mihalas 1984).  Eq.~\ref{L_eq} has several shortcomings in the region above
the photosphere.  First, it implies an Eddington factor of $f_E=1/3$, which is
not correct in the optically thin outer region where $f_E$ approaches 0.4--0.5
(Feuchtinger \& Dorfi 1994, Fig.~3).  Second, in this regime, the opacities
should be higher than those given by the Rosseland mean (Alexander \& Ferguson
1994).  Finally, for the computation of the periods and growth rates it would
be more appropriate to linearize the radiation hydrodynamics equations rather
than the equilibrium diffusion equation.  The magnitude of the errors
introduced by the first effect can easily be estimated.

One can approximate the effect of the Eddington factor in Eq.~\ref{L_eq}. by
replacing $3\kappa$ by $\kappa/f_E$.  The Eddington factor also appears in the
radiation pressure which becomes $p_{\rm rad}\rightarrow 3f_E \th p_{\rm rad}$.
We can disregard sphericity effects in the momentum equation contained in the
term $(p_{\rm rad}-f_E\th e_{\rm rad})/r$.

To simulate the effect of an increasing $f_E$ we assume that for $T <$ \Teff,
$f_E$ increases smoothly from 1/3 to 1/2 with decreasing temperature.  The
results confirm our hunch that an increase of $f_E$ in the very outer region
has very little influence on the growth rates and almost none on the
periods. More importantly it has no differential sensitivity to luminosity.

\centerline{\vbox{\epsfxsize=9cm\epsfbox{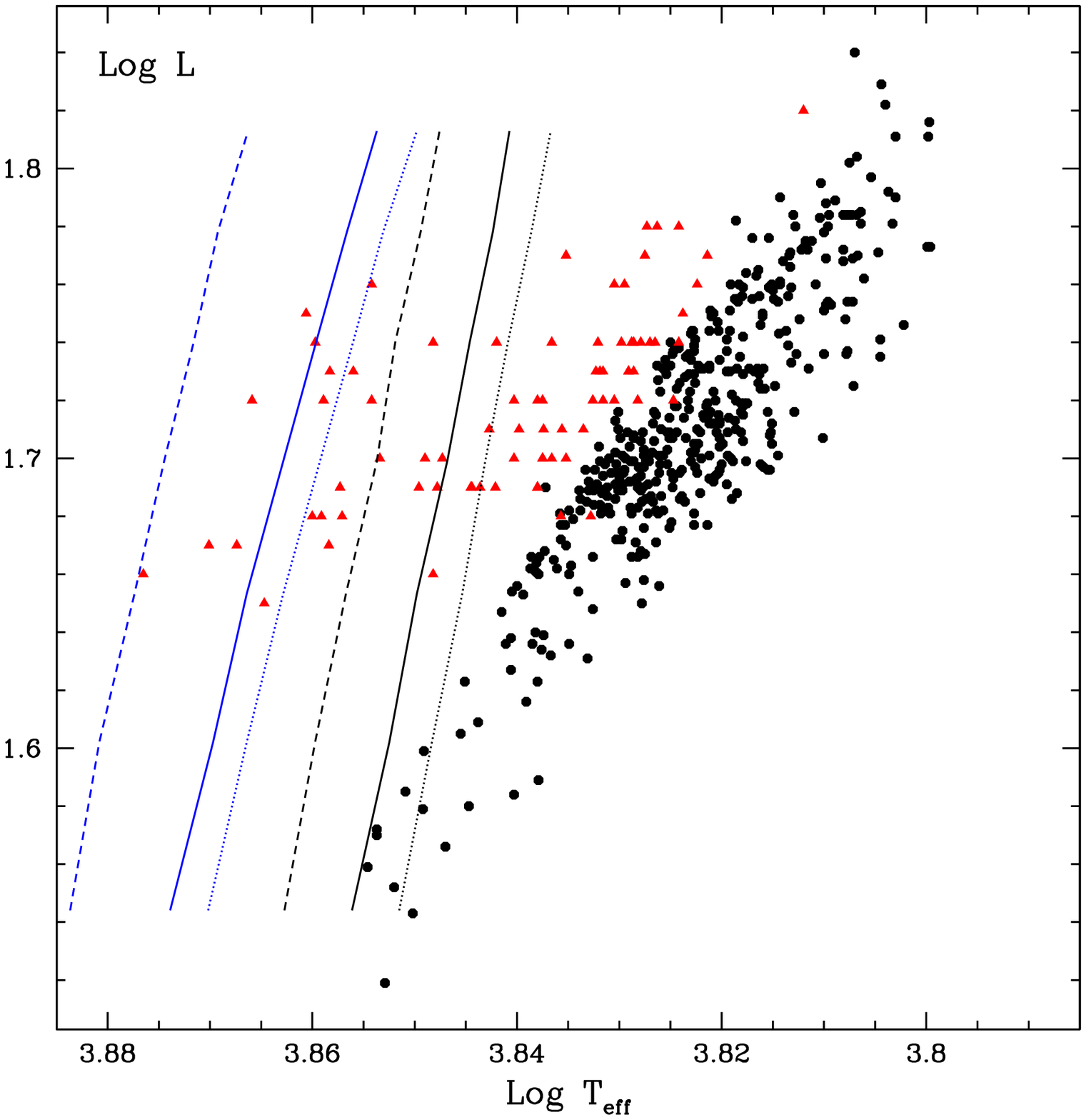}}}
\noindent{\small
 Fig.~2 :
 \LogL-\LTeff\ plot:
 RRab (filled circles) and RRc (triangles) stars of Jurcsik;
 Radiative blue edges for M=0.65;
 solid line:   X=0.75 Z=0.001,
      dotted line:  X=0.75 Z=0.004,
      dashed line:  X=0.70 Z=0.001.
 }
 \label{fig2}

 \centerline{\vbox{\epsfxsize=9cm\epsfbox{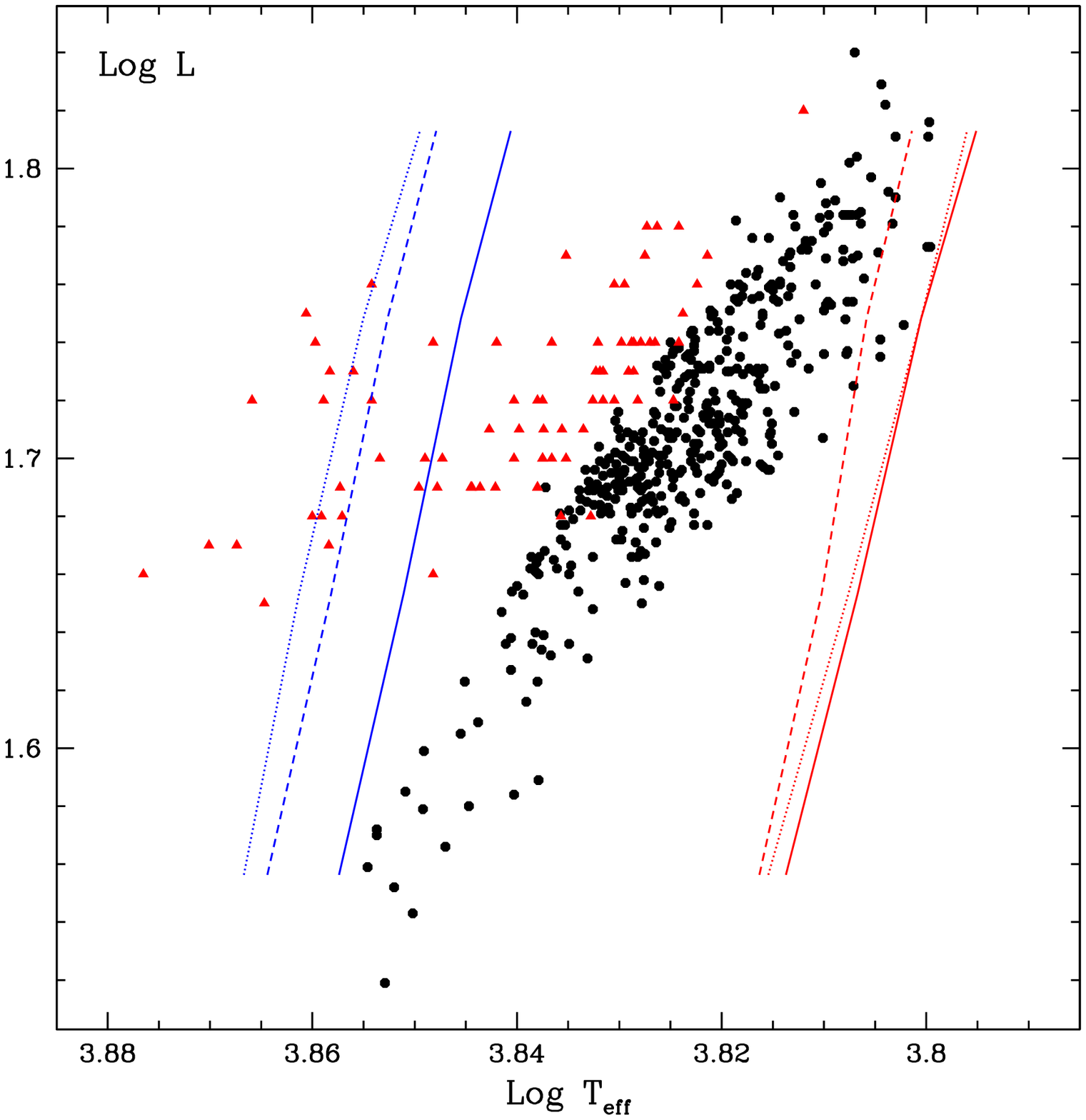}}}
 \noindent{\small
 Fig.~3 :
 \LogL-\LTeff\ plot:
 RRab (filled circles) and RRc (triangles) stars:
  Convective models:  Solid, dotted and dashed lines are for three different
 combinations of the $\alpha$ parameters in the convective model equations,
 chosen to give a reasonable width for the instability strip.
 }
 \label{fig3}
 \vspace{5mm}

\begin{figure*}
\centerline{{\vbox{\epsfxsize=9cm\epsfbox{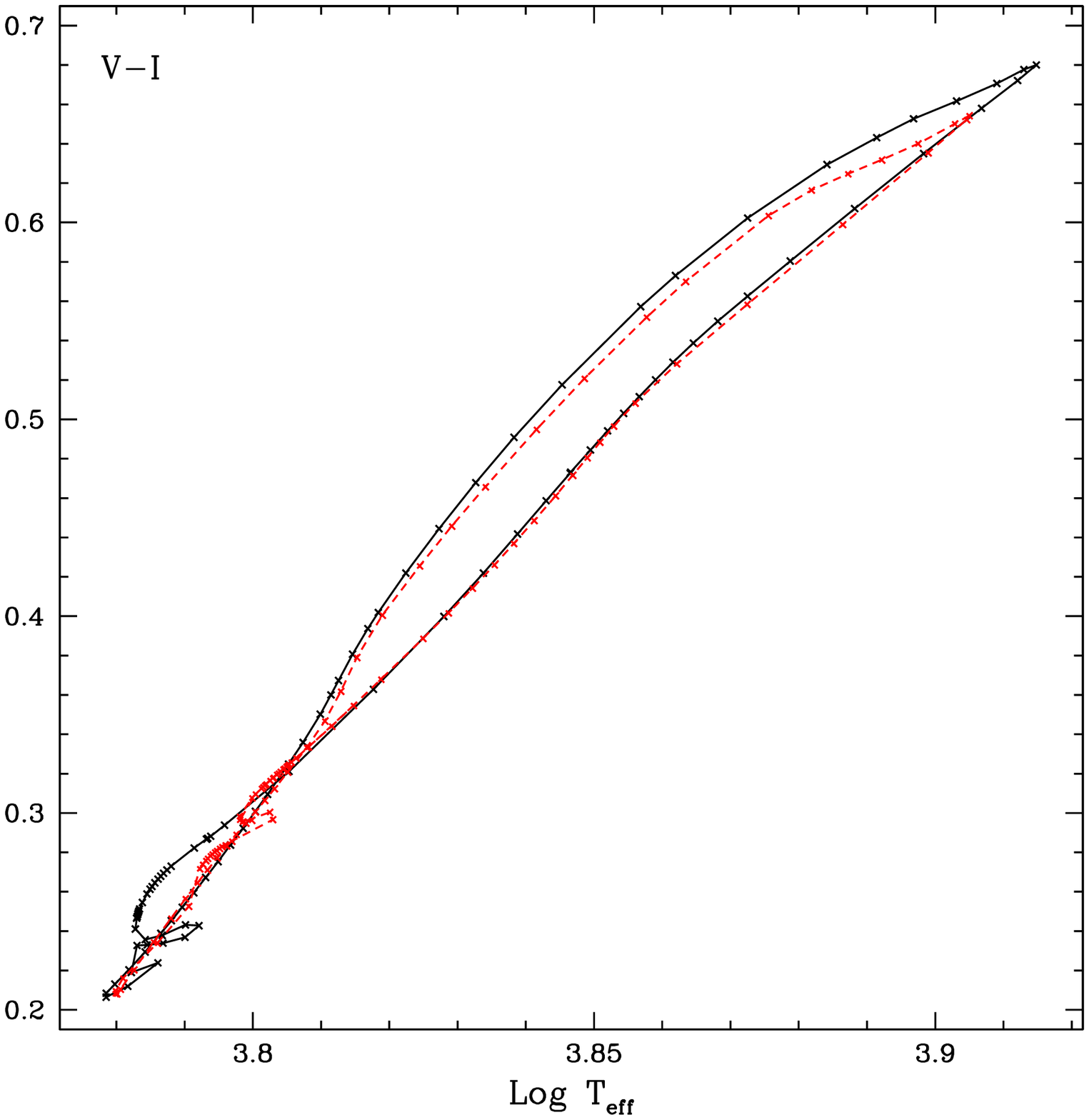}}}
            {\vbox{\epsfxsize=9cm\epsfbox{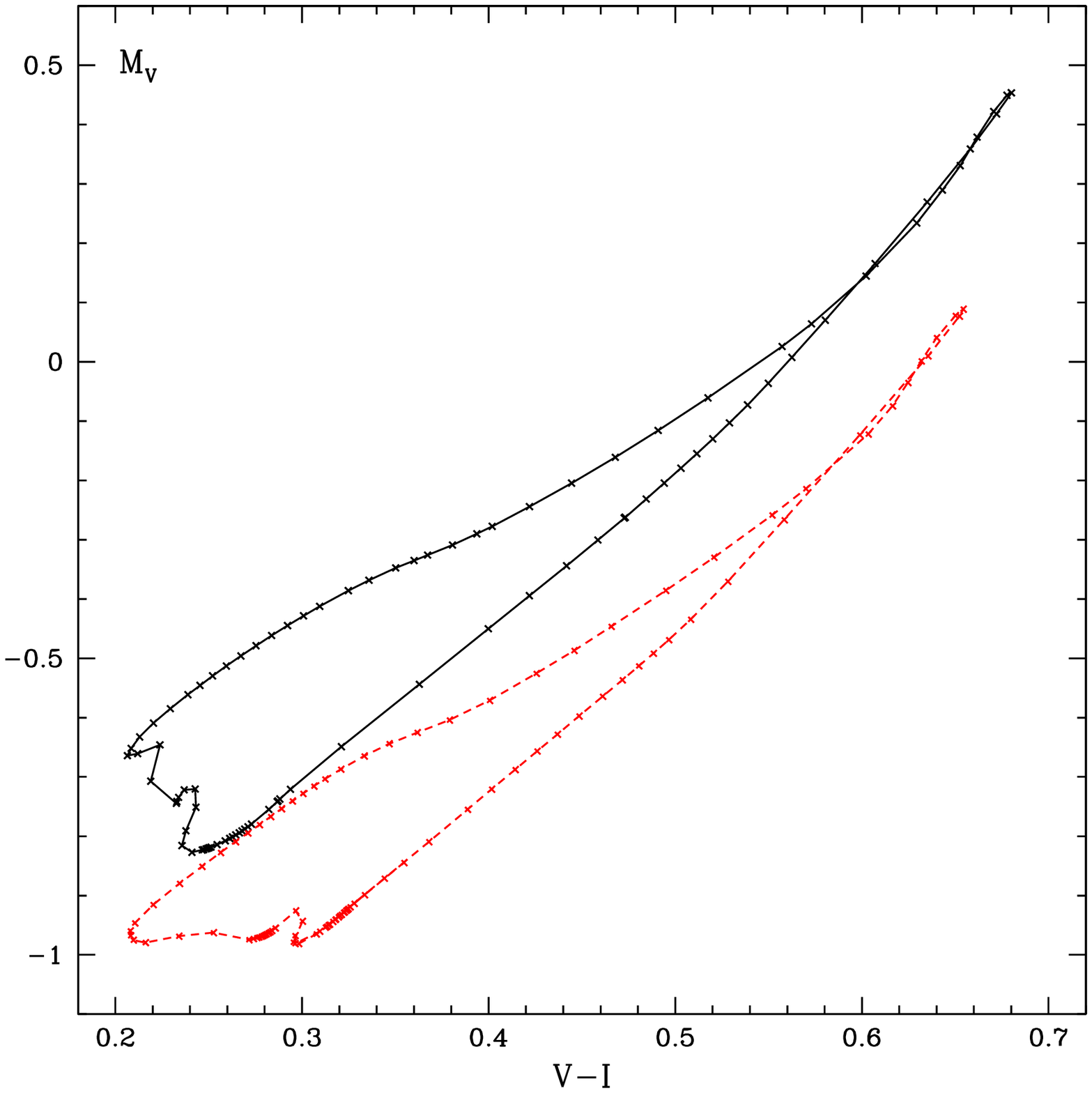}}}}
\vspace{0.2cm} \noindent{\small
 Fig.~4 :
 Left: \Teff-(V-I) plot and right: (V-I)-M$_V$ plot:
 Theoretical color variations for two nonlinear full amplitude RRab models with
 different luminosities of L = 40 \Lo (dashed line) and L = 52 \Lo (full
 line). Crosses indicate the time steps of the nonlinear
 calculations. See text for details.}
 \label{fig4}
\vskip 0.5cm
\end{figure*}

We have artificially, and somewhat arbitrarily, increased the opacity from the
vicinity of \Teff\ outward.  We have not found an appreciable differential
effect with \Teff, and have therefore not pursued this avenue with more
detailed modelling.

Finally, the insensitivity of the slope to $f_E$ makes us believe that the
linearization of the radiation hydrodynamics equations instead of the common
equilibrium diffusion equation (requiring a serious coding effort) would also
make little difference.

We conclude that the discrepancy between the models and the observations 
is not due to an inadequate treatment of the radiation transport.

 \subsection{Composition}

Could compositional make-up be more important than expected?  We have computed
a number of RR Lyrae models with various combinations of Y and Z, but find that
the effect on $\Xi$ is negligible.  Next we have artificially increased the
abundances of the easily ionizable light elements such as Mg and Na in the OPAL
opacities.  Again this has such a negligible effect on $\Xi$ that the inclusion
of this data would unnecessarily clutter Figure~2.

An inhomogeneous composition is unlikely to exist in RR Lyrae envelopes because
of convection, but even if existed our tests with various changes in
composition make us doubt that it would resolve the discrepancy.

 \subsection{Rotation}

Could rotation be responsible for the discrepancy?  In order to estimate the
magnitude of the effect of rotation we have included a spherical
pseudo-centrifugal acceleration $\omega^2\th r$ in the equilibrium model and in
the computation of the periods and growth rates.  We find that a rather short
rotation period, of order of a few days, would be necessary to have an impact.
Furthermore the disagreement is worsened by rotation because the slope
steepens.

We conclude that moderate rotation rates cannot be the cause of the
discrepancy.

\subsection{Evolutionary Effects}

 The blueward moving evolutionary paths turn around at some \Teff\ and then
move upward (\eg Dorman 1992, Lee, Demarque \& Zinn 1990).  If this happened
inside the linear instability strip, then the leftmost blue edge, that of the
overtone mode, would be defined by the topology of these paths, rather than by
the pulsational stability of the models.  However, the fact that the Jurcsik
blue and red edges of the fundamental and the overtone are all essentially
parallel eliminates this possibility as an explanation for the shallow observed
slope.  Moreover, for some of the clusters, full photometry of the horizontal
branch exists, and there is no indication of any gap next to the RR Lyrae
region.  Clearly there are stars next to the empirical red edge for which the
amplitude (if not zero) is less than the observational limit.

\subsection{Conclusion}

The theoretical $\Xi$ slope is very robust with respect to the uncertainties or
approximations inherent in our code, but is in disagreement with the
Jurcsik empirical relations.  We have therefore been led to reexamine
the latter for their robustness, especially since they also make use of some
theory, \viz color to temperature transformation.

\vskip 20pt

 \section{Discussion of the Empirical Relations}

Because of the strong disagreement between Jurcsik's processed observational
data and the predictions from something as basic as radiative models, it is
worth while to reexamine some of the most uncertain points in the reduction of
the observational data.

The observational material consists of RRab periods and lightcurves.  KJ96,
JK97 and J98 deduce the color, the visual magnitude M$_v$ and the metallicity
Z ([Fe/H] in the form of linear fully empirical relations


\begin{eqnarray}
 \pmatrix{M_v    \cr
           V-K    \cr
          [Fe/H] \cr}
      &\equiv&  c_0 + A     \cdot   \pmatrix{P        \cr
                                             A_1      \cr
                                             \phi_{31}\cr
                                             \phi_{41}\cr}
\label{eq_empir}
\end{eqnarray}

The connection with \LTeff\ and with \LogL\ was then made through static
envelope models of Kurucz (1993, \cf J98 for details).  This is the only
theoretical input into the otherwise empirical relations.

\begin{equation}
 \pmatrix{\Log L         \cr
          \L T_{ef\!f} \cr}
       = c_1 + C
                             \cdot  \pmatrix{M_v\cr
                                             V-K \cr
                                            [Fe/H] \cr}
       = c_2 + C\cdot A
                             \cdot  \pmatrix{P       \cr
                                             A_1      \cr
                                             \phi_{31}\cr
                                             \phi_{41}\cr}
\label{eq_kur}
\end{equation}

\begin{figure*}
\centerline{{\vbox{\epsfxsize=9cm\epsfbox{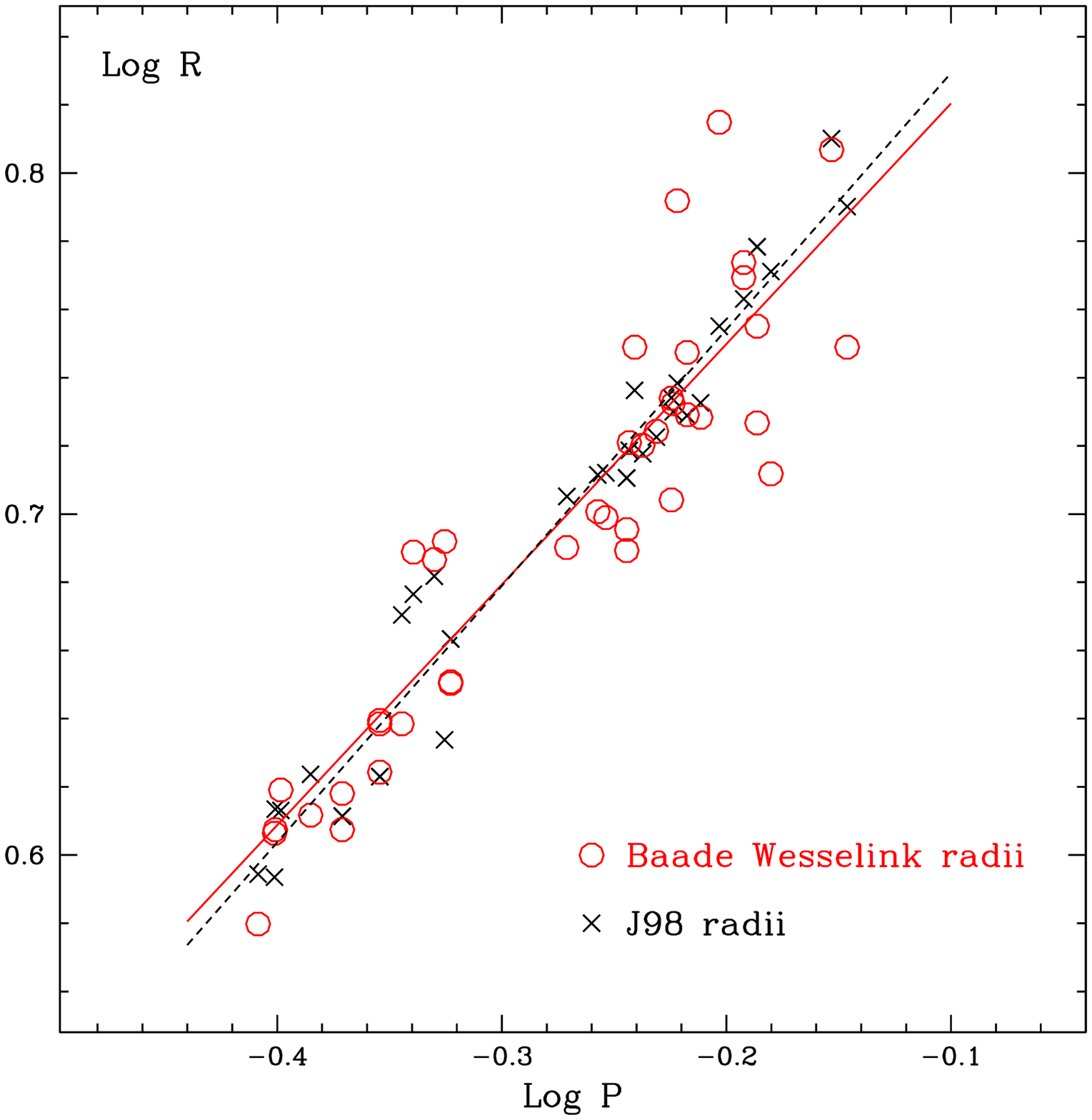}}}
            {\vbox{\epsfxsize=9cm\epsfbox{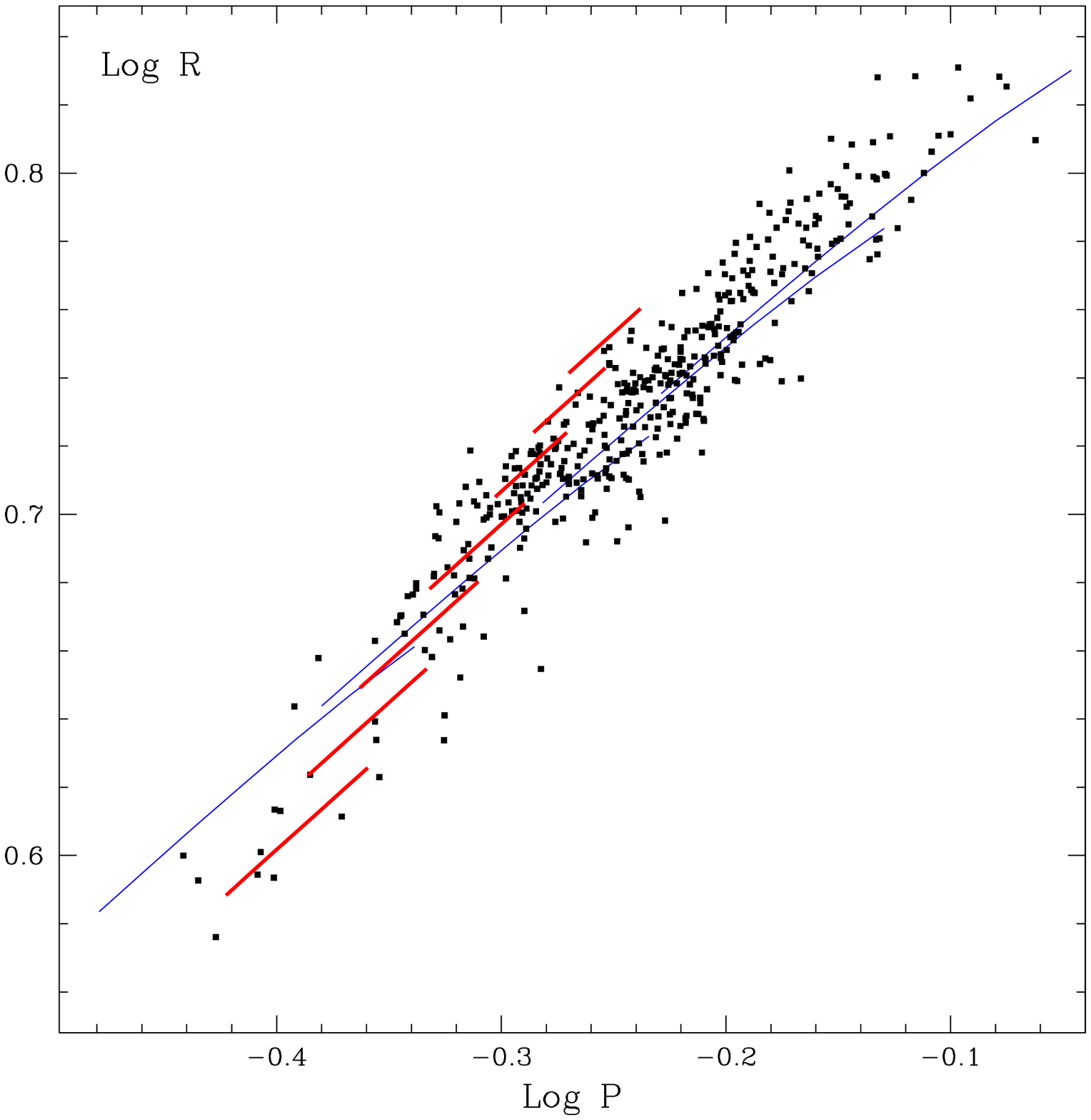}}}}
\vspace{0.2cm}
\noindent{\small
 Fig.~5 : Left: RRab Period -- Radius plot;
 {\it open circles}: Baade-Wesselink radii, and
 {\it crosses}: same stars from J98;
 The solid and dashed lines are linear regressions through the Baade-Wesselink
and the Jurcsik data, respectively.
 Right: 
 {\it thick lines}: radiative models with Jurcsik L, M, Z relation;
 {\it thin lines}: radiative models with constant mass and luminosity;
 {\it filled squares}: same stars from J98; 
}
 \label{fig5}
\vskip 1cm
\end{figure*}

\vspace{0.5cm}

\subsection{Color -- \Teff\ Transformation}

J98 uses the {\it static} envelope calculations of Kurucz's ATLAS code (for
details cf. J98), to make the transformation from average color to average
\Teff.  The \Teff\ however varies considerably during one pulsation cycle,
compared to the width of the instability strip, and the question arises whether
the correct mean color and mean temperature are still related through the
static Kurucz relations.  To check the validity of this assumption we have
computed the spectra of two full amplitude RR Lyrae models.

We investigate the theoretical color variation for two RR Lyrae models with
different luminosities (L = 40 \Lo\ and L = 52 \Lo, both for M = 0.65 \Mo,
\Teff = 6500 K, X = 0.76 and Z=0.001).  Nonlinear full amplitude models are
taken from Feuchtinger (1999) and the color variations in V and I are computed
as outlined in Dorfi \& Feuchtinger (1999).  The results can be inferred from
Fig.~\ref{fig4} which displays the color difference V -- I as a function of the
effective temperature (left panel) and the corresponding color-magnitude
diagram (right panel), both over one oscillation cycle.  Full and dotted lines
refer to the L = 40 \Lo~and L = 52 \Lo\ models, respectively.  The differential
effect with L is seen to be small.  Dynamical atmospheric effects thus cannot
account for the discrepancy.

There remains the question of the consistency of the static Kurucz atmospheres
with our frequency dependent calculation.  For that purpose we have checked the
color -- temperature transformation against the Kurucz tables at selected
points on the pulsation cycle.  The agreement is quite satisfactory, and any
discrepancies are irrelevant for our purpose here.

\vspace{0.9cm}

\subsection{RR Lyrae Period - Radius Relation}

The radii of RR Lyrae stars can be obtained independently either from a
Baade-Wesselink approach or from the J98 empirical L and \Teff.  In
Fig.~\ref{fig5}, on the left, we have plotted as open circles the
Baade-Wesselink radii of the 15 RRab stars of Jones \etal (1992) (we have
omitted RR~Leo).  Superposed as crosses are the radii for the same stars
derived from the J98 \th L and \Teff.  The agreement is remarkably good, but
worsens for the longer periods.  The solid and dashed lines represent linear
regression lines for the BW and for the J98 period -- radius relations.

We have also derived the radii of the additional J98 stars that were not in the
Jones sample and show them as small dots in the righthand figure.  The thin
lines represent constant mass and constant luminosity radiative models, and the
thick lines radiative models that obey the J98 \th L, M, Z relations
Since the radiative models do not provide a red edge, the upward extent of the
thin or thick lines is not significant.  The latter models are seen to give
better agreement with the swarm of the J98 empirical radii.

We conclude that the agreement between the empirical J98 and Jones'
Baade-Wesselink radii is remarkably good, and so is the agreement that can be
achieved with theoretical models.  A consideration of the period -- luminosity
relations therefore does not give us any clues as to the origin of the slope
discrepancy.

\vfill\eject

\section{Conclusions}

We have shown that the Jurcsik relations lead to RR Lyrae radii that are in
good agreement with both Baade-Wesselink radii and with theoretical radii.  On
the other hand, there exists a strong discrepancy between the slope of the
theoretical \LogL -- \LTeff\ relation and the slope of the empirical Jurcsik
relation.  

We have reviewed the physical and numerical uncertainties that enter the
theoretical calculations.  The discrepancy exists already at the level of
purely radiative modelling.  We have further shown that our model equations for
turbulent convection do not alter the slope of the \LogL--\LTeff\ relation
despite the large number (8) of adjustable parameters ($\alpha$'s) that we have
at our disposal.  Consequently, apart from the uncertainties inherent in a 1D
recipe for turbulent convection, it appears that the discrepancy is not caused
by a deficiency of the theoretical models.  We note that the same slope
discrepancy is implicit in older and more recent independent calculations
(Tuggle \& Iben 1972, Bono \etal 1997).
  
The derivation of the empirical relations makes use of a {\it static} Kurucz
color - \Teff\ transformation.  By computing the behavior of color versus
temperature over the pulsation cycle we have shown that the use of static
envelopes is in fact a very good approximation, and therefore cannot be the
culprit.

Finally, the shape of the evolutionary tracks, through their potential
avoidance of certain regions of the instability strip cannot be responsible
either.  

At this time the origin of this disturbing discrepancy constitutes an unsolved
puzzle.  It remains to be seen whether an improved treatment of turbulent
convection or more complete observations will resolve the difficulty.

 \section{Acknowledgements} The authors would like to thank Johanna Jurcsik for
kindly providing us with copies of her data sets.  They also acknowledge
fruitful discussions with Johanna Jurcsik and G\'eza Kov\'acs.  This work has
been supported by NSF (AST 95-28338 and AST 98-19608) and the Hungarian OTKA
(T-026031).

\end{document}